\documentclass[prd,preprintnumbers,nofootinbib,onecolumn]{revtex4} 

\usepackage{graphicx}
\usepackage{dcolumn}
\usepackage{bm}
\usepackage{latexsym}
\usepackage{amsfonts}
\usepackage{amssymb}
\usepackage{amsmath}

\begin{document}

\preprint{IPMU11-0061, YITP-11-50}

\title{Nonlinear superhorizon perturbations in Ho\v{r}ava-Lifshitz gravity}

\author{Keisuke Izumi}
\email{izumi@yukawa.kyoto-u.ac.jp}
\affiliation{
Yukawa Institute for Theoretical Physics, 
Kyoto University, Kyoto 606-8502, Japan}
\author{Shinji Mukohyama}
\email{shinji.mukohyama@ipmu.jp}
\affiliation{
 IPMU, The University of Tokyo, Kashiwa, Chiba 277-8582, Japan}

\date{\today}

\begin{abstract}
 We perform a fully nonlinear analysis of superhorizon perturbation in
 Ho\v{r}ava-Lifshitz gravity, based on the gradient expansion method. We
 present a concrete expression for the solution of gravity equations up
 to the second order in the gradient expansion, and prove that the
 solution can be extended to any order. The result provides yet another
 example for analogue of the Vainshtein effect: the nonlinear solution
 is regular in the limit $\lambda\to 1$ and recovers general relativity
 coupled to dark matter at low energy.
 Finally, we propose a definition of nonlinear
 curvature perturbation ${\cal R}$ in Ho\v{r}ava-Lifshitz gravity and
 show that it is conserved up to the first order in the gradient
 expansion. 
\end{abstract}

\maketitle

\section{Introduction}

The new theory of gravitation proposed recently by
Ho\v{r}ava~\cite{Horava:2009uw,Mukohyama:2010xz} is expected to be renormalizable and
unitary. For this reason, it has been attracting significant amount of
attention~\cite{Charmousis:2009tc,Koyama:2009hc,Mukohyama:2009mz,Mukohyama:2009gg}. 
The theory is power-counting renormalizable because of the
anisotropic scaling in the ultraviolet (UV), 
\begin{eqnarray}
 t \to b^z t, \qquad \Vec x \to b \Vec x, 
  \label{scalling}
\end{eqnarray}
with the dynamical critical exponent $z\ge 3$. Since this scaling is
called Lifshitz scaling, the theory is often called Ho\v{r}ava-Lifshitz
gravity.

Because of the anisotropic scaling (\ref{scalling}), the time and the
space in this theory must be treated separately. Thus, we must abandon
the $4$-dimensional diffeomorphism invariance. Instead, the fundamental
symmetry of the theory is the invariance under the so-called
foliation-preserving diffeomorphism: 
\begin{equation}
 t \to t'(t), \quad \vec{x}\to\vec{x}'(t,\vec{x}), \label{foliation}
\end{equation}
which preserves the way the spacetime is foliated by constant-time
hypersurfaces. Since this symmetry is smaller than the counterpart of
general relativity i.e. the $4$-dimensional diffeomorphism invariance,
the structure of the action is less restrictive and allows more
parameters. For example, the kinetic part of the action is a linear
combination of $K^2$ and $K^{ij}K_{ij}$ with arbitrary
coefficients, where $K_{ij}$ is the extrinsic curvature of the
constant-time hypersurface and $K=K^i_{\ i}$. Hence, the kinetic part
can be written as 
\begin{eqnarray}
I_{kin}=\frac{M_{Pl}^2}{2}\int Ndt \, \sqrt{g}d^3\vec{x}
 \left(K^{ij} K_{ij} - \lambda K^2\right),
\end{eqnarray}
where $M_{Pl}$ is the reduced Planck mass and $\lambda$ is an arbitrary
parameter. While in general relativity the value of $\lambda$ is fixed
to unity due to the 4-dimensional diffeomorphism invariance, 
in Ho\v{r}ava-Lifshitz gravity any value of $\lambda$ is consistent
with the foliation-preserving diffeomorphism invariance.

In Ho\v{r}ava-Lifshitz gravity, in addition to the usual tensor
gravitons, there is an extra physical degree of freedom called the
scalar graviton~\cite{Horava:2009uw}. In order for the scalar graviton not to be a
ghost, the regime $1/3<\lambda<1$ should be excluded~\cite{Charmousis:2009tc}. 
Outside this
forbidden interval, the scalar graviton has negative sound speed 
squared. Therefore, in order for the theory to be observationally
viable, we need to impose the condition under which the associated
long-distance instability does not show up~\cite{Mukohyama:2010xz}. The
condition essentially says that $\lambda$ must be sufficiently close to
$1$ in the infrared (IR), and should be considered as a phenomenological
constraint on properties of the renormalization group (RG) flow. 
Since the value of $\lambda$ continuously changes by the RG flow, 
only the regime $\lambda>1$ is allowed.

When $\lambda$ is very close to $1$, the scalar graviton gets strongly
coupled~\cite{Koyama:2009hc}. If we adopt the usual metric perturbation method then 
we find that higher order terms in the time kinetic part of the action for the
scalar graviton become larger. This indicates the breakdown of the
perturbative expansion in the scalar graviton sector. Note that this
does not necessarily imply the loss of predictability since all
coefficients of infinite number of terms can be written in terms of
finite parameters in the action if the theory is
renormalizable. However, because of the breakdown of the perturbative
expansion, we need to employ a more or less non-perturbative method to 
analyze the fate of the scalar graviton in the limit $\lambda\to 1$. 
Such an analysis was performed in \cite{Mukohyama:2010xz} for
spherically symmetric, static, vacuum configurations and it was shown
that the limit is continuous and recovers general relativity. This may
be considered as an analogue of the Vainshtein
effect~\cite{Vainshtein:1972sx}.

The main purpose of this paper is to provide yet another
example of the analogue of the Vainshtein effect in Ho\v{r}ava-Lifshitz
gravity. For this purpose, we perform a fully nonlinear analysis of
superhorizon cosmological perturbation, adopting the so-called gradient
expansion method~\cite{Salopek:1990jq}. The result is obviously
continuous in the limit $\lambda\to 1$ and recovers general relativity
coupled to dark matter.

We also propose a definition of nonlinear curvature perturbation 
${\cal R}$ in Ho\v{r}ava-Lifshitz gravity and show that it is conserved
up to the first order in the gradient expansion.

The paper is organized as follows. In \S~\ref{basic}, we briefly
review the basic equations in Ho\v{r}ava-Lifshitz gravity. In
\S~\ref{solutions}, we introduce the gradient expansion method in
this theory and present the solution to equations of motion. In
\S~\ref{CP}, we proposed a definition of nonlinear curvature
perturbation ${\cal R}$ and show that it is conserved up to the first
order in the gradient expansion. \S~\ref{sec:summary} is devoted to a
summary of this paper. In appendix \ref{proof}, we prove that the
solution obtained in \S~\ref{solutions} satisfies the momentum
constraint in any order of the gradient expansion.

\section{Basic equations}
\label{basic}

In this section we review the basic equations of Ho\v{r}ava-Lifshitz
gravity, following the notation in \cite{Mukohyama:2010xz}, and
reformulate them in a way suitable for gradient expansion. Basic
quantities of Ho\v{r}ava-Lifshitz gravity are the lapse $N(t)$, the
shift $N^i(t,\vec{x})$ and the $3$-dimensional spatial metric
$g_{ij}(t,\vec{x})$. Combining these quantities, we can construct
$4$-dimensional spacetime metric in the ADM form as 
\begin{eqnarray}
ds^2= -N^2dt^2+g_{ij}(dx^i+N^idt)(dx^j+N^jdt).
\label{metric}
\end{eqnarray}
The fundamental symmetry of the theory is the invariance under the so
called foliation preserving diffeomorphism:
\begin{equation}
 t \to t'(t), \quad \vec{x}\to\vec{x}'(t,\vec{x}),
\end{equation}
which preserves the way the spacetime is foliated by constant-time
hypersurfaces.

By requiring invariance under spatial parity and time reflection, the
gravitational action is specified as
\begin{eqnarray}
I_g=\frac{M_{Pl}^2}{2}\int Ndt \, \sqrt{g}d^3\vec{x}
 \left(K_{ij} K^{ij} - \lambda K^2 -2\Lambda+R+L_{z>1}\right) ,
\label{eq:action}
\end{eqnarray}
where $\sqrt{g}$ is the determinant of $g_{ij}$, $K_{ij}$ is the
extrinsic curvature defined as
\begin{eqnarray}
K_{ij}=\frac{1}{2N}\left(
\partial_t g_{ij}-D_i N_j-D_j N_i
\right) ,
\end{eqnarray}
$K$ ($=g^{ij}K_{ij}$) is the trace of $K_{ij}$, $D_i$ is the spatial
covariant derivative compatible with $g_{ij}$, and $R$ is the Ricci
scalar constructed from $g_{ij}$. To lower and raise an index, $g_{ij}$
and its inverse $g^{ij}$ are used. We impose the so called
projectability condition, namely we require that the lapse function
should depend only on time. In order to realize the power-counting
renormalizability, the higher curvature Lagrangian $L_{z>1}$ should
include up to sixth or higher spatial derivatives. For our analysis in
this paper, we do not need to specify the concrete form of
$L_{z>1}$. Moreover, for simplicity we shall not include matter action
and analyze the pure gravity described by the action (\ref{eq:action}).

The equation of motion for $g_{ij}$ is ${\cal E}_{g ij}=0$, where 
\begin{eqnarray}
 {\cal E}_{g ij} & \equiv &
  g_{ik}g_{jl}\frac{2}{N\sqrt{g}}
  \frac{\delta I_g}{\delta g_{kl}}\nonumber\\
& = & M_{Pl}^2
  \left[
  -\frac{1}{N}(\partial_t-N^kD_k)p_{ij}
  + \frac{1}{N}(p_{ik}D_jN^k+p_{jk}D_iN^k)
  \right.\nonumber\\
 & & \left.
  - Kp_{ij} + 2K_i^kp_{kj}
      + \frac{1}{2}g_{ij}K^{kl}p_{kl}-\Lambda g_{ij} 
      - G_{ij}\right]
 + {\cal E}_{z>1 ij}.
\end{eqnarray} 
Here, $p_{ij}\equiv K_{ij}-\lambda Kg_{ij}$, ${\cal E}_{z>1 ij}$ is the
contribution from $L_{z>1}$ and $G_{ij}$ is Einstein tensor of
$g_{ij}$. The trace part and traceless part of this equation are,
respectively, 
\begin{equation}
 (3\lambda-1)\left(\partial_{\perp}K+\frac{1}{2}K^2\right)
  + \frac{3}{2}A^i_{\ j}A^j_{\ i} + Z = 0,
  \label{eq:Tr}
\end{equation}
and 
\begin{equation}
 \partial_{\perp}A^i_{\ j} + KA^i_{\ j}
  + \frac{1}{N}(A^k_{\ j}\partial_kN^i-A^i_{\ k}\partial_jN^k)
  -\left(Z^i_{\ j}-\frac{1}{3}Z\delta^i_j\right) = 0,
\label{eq:Trless}
\end{equation}
where 
\begin{equation}
 A^i_{\ j} \equiv K^i_{\ j} - \frac{1}{3}K\delta^i_j
\end{equation}
is the traceless part of $K^i_{\ j}$, 
\begin{equation}
 \partial_{\perp} \equiv \frac{1}{N}(\partial_t-N^k\partial_k),
\end{equation}
and
\begin{equation}
 Z^i_{\ j} \equiv -\Lambda\delta^i_j - G^i_{\ j}
  + M_{Pl}^{-2} g^{ik}{\cal E}_{z>1 kj}, \quad
  Z = Z^i_{\ i} = -3\Lambda + \frac{1}{2}R
  + M_{Pl}^{-2}g^{ij}{\cal E}_{z>1 ij}.
\label{eq:defZ}
\end{equation}
The foliation preserving diffeomorphism includes the three dimensional
spatial diffeomorphism as a part of it. As a result, $Z^i_{\ j}$
satisfies the generalized Bianchi identity,
\begin{eqnarray}
D_j Z^j_{\ i}=0.\label{eq:DZ}
\end{eqnarray}

For convenience, we decompose the spatial metric and the extrinsic curvature as
\begin{eqnarray}
g_{ij} & = & a^2(t) e^{2\zeta(t,\vec{x})} \gamma_{ij}(t,\vec{x}), \\
K^i_{\ j} & = & \frac{1}{3} K(t,\vec{x}) \delta^i_{\ j}
 +A^i_{\ j}(t,\vec{x}),
 \label{eq:decomposition}
\end{eqnarray}
where we define $\zeta(t,\vec{x})$ so that $\det\gamma =1$, and $a(t)$
is defined in eq.(\ref{defa}) later. The trace part and the traceless
part of the definition of the extrinsic curvature lead, respectively, to
\begin{equation}
\partial_{\perp} \zeta+ \frac{\partial_t a}{Na}
 = \frac{1}{3}\left(K +\partial_iN^i\right),
\label{eq:TrK}
\end{equation}
and
\begin{equation}
 \partial_{\perp}\gamma_{ij}
  = 2\gamma_{ik}A^k_{\ j}
  + \frac{1}{N}
  \left(\gamma_{jk}\partial_iN^k
  + \gamma_{ik}\partial_jN^k
  -\frac{2}{3}\gamma_{ij}\partial_kN^k
       \right).
\label{eq:TrlessK}
\end{equation}

The momentum constraint, i.e. the equation of motion for $N^i$, is
\begin{equation}
 D_jK^j_{\ i}-\lambda \partial_iK = 0. 
\end{equation}
According to the decomposition (\ref{eq:decomposition}), the momentum
constraint is rewritten as
\begin{equation}
\partial_jA^j_{\ i} + 3A^j_{\ i}\partial_j\zeta 
 - \frac{1}{2}A^j_{\ l}\gamma^{lk}\partial_{i}\gamma_{jk}
 - \frac{1}{3}\left(3\lambda-1\right)\partial_iK = 0.
\label{eq:momentum-constraint}
\end{equation}

As a consistency check, it is instructive to calculate time derivatives of
$A^i_{\ i}$, $\ln\det\gamma$,
$\gamma_{ij}-\gamma_{ji}$ and
$\gamma_{ik}A^k_{\ j}-\gamma_{jk}A^k_{\ i}$ 
without using the fact that they actually vanish. The results are
\begin{eqnarray}
\partial_{\perp} A^i_{\ i} & = & -KA^i_{\ i}, \nonumber\\
 \partial_{\perp}(\ln\det\gamma) & = & 
  2 A^i_{\ i}, \nonumber\\
 \partial_{\perp}(\gamma_{ij}-\gamma_{ji})
  & = & 2(\gamma_{ik}A^k_{\ j}-\gamma_{jk}A^k_{\ i}), 
  \nonumber\\
 \partial_{\perp}(\gamma_{ik}A^k_{\ j}-\gamma_{jk}A^k_{\ i})
  & = & -K(\gamma_{ik}A^k_{\ j}-\gamma_{jk}A^k_{\ i})
  + 2(\gamma_{il}A^l_{\ k}-\gamma_{kl}A^l_{\ i})A^k_{\ j}
  - 2(\gamma_{jl}A^l_{\ k}-\gamma_{kl}A^l_{\ j})A^k_{\ i}
  \nonumber\\
 & & 
  + 2(\gamma_{kl}-\gamma_{lk})A^k_{\ i}A^l_{\ j}
  +\frac{1}{N}(\gamma_{ik}A^k_{\ l}-\gamma_{lk}A^k_{\ i})
  \partial_jN^l
  -\frac{1}{N}(\gamma_{jk}A^k_{\ l}-\gamma_{lk}A^k_{\ j})
  \partial_iN^l \nonumber\\
 & &
  - \frac{2}{3}(\gamma_{ik}A^k_{\ j}-\gamma_{jk}A^k_{\ i})
  \partial_lN^l
  - \frac{1}{3}Z(\gamma_{ij}-\gamma_{ji}).
\end{eqnarray}
The right hand side of each equation vanishes when $A^i_{\ i}$,
$\ln\det\gamma$, $\gamma_{ij}-\gamma_{ji}$ and 
$\gamma_{ik}A^k_{\ j}-\gamma_{jk}A^k_{\ i}$ vanish. 
Therefore, the evolution equations we have derived are consistent with
vanishing $A^i_{\ i}$, $\ln\det\gamma$, 
$\gamma_{ij}-\gamma_{ji}$ and
$\gamma_{ik}A^k_{\ j}-\gamma_{jk}A^k_{\ i}$.

\section{Gradient expansion}
\label{solutions}

The gradient expansion is the method to analyze the full non-linear
dynamics at large scale. In the gradient expansion, we consider
nonlinear perturbation around a flat Friedmann-Robertson-Walker
background and suppose that the characteristic spatial scale $L$ of the
perturbation is much larger than the Hubble horizon scale $1/H$. To make
the argument transparent, we introduce a small parameter $\epsilon$
defined by $\epsilon\sim 1/(HL)$ and expand all relevant quantities and 
equations with respect to $\epsilon$. For example, a spatial derivative
acted on a relevant quantity raises the order of $\epsilon$ and thus is
counted as $O(\epsilon)$. We then solve the equations order by order in
gradient expansion. 

\subsection{Gauge fixing}

The foliation preserving diffeomorphism invariance is, like all other
gauge symmetries, redundancy of descriptions. In order to extract
physical quantities and statements, we thus need to eliminate gauge
freedom by imposing appropriate gauge condition. In this paper we adopt
the synchronous gauge, or the Gaussian normal coordinate system, by
setting the lapse to unity and the shift to zero. 
\begin{eqnarray}
N=1, \ N^i= 0.\label{eq:gaugefix}
\end{eqnarray}
This fixes the time coordinates but does not completely fix the spatial
coordinates. There still remains gauge freedom of time-independent
spatial diffeomorphism, corresponding to the change of coordinates on
the initial constant-time hypersurface. This residual gauge degree of
freedom will be discussed later.

In this gauge our basic equations (\ref{eq:Tr}), (\ref{eq:Trless}),
(\ref{eq:TrK}) and (\ref{eq:TrlessK}) are simplified as
\begin{eqnarray}
(3\lambda-1) \partial_t K & = &
 -\frac{1}{2}(3\lambda-1)K^2 -\frac{3}{2}A^i_{\ j}A^j_{\ i} - Z,
 \label{eq:K}\\
\partial_t A^i_{\ j} & = & 
 -KA^i_{\ j}+Z^i_{\ j}-\frac{1}{3}Z\delta^i_{\ j},\label{eq:A}\\
\partial_t \zeta & = & 
 -\frac{\partial_t a}{a}+\frac{1}{3}K,\label{eq:psi}\\
\partial_t \gamma_{ij} & = &
 2\gamma_{ik}A^k_{\ j}.\label{eq:evogama}
\end{eqnarray}
Hereafter we assume that $\lambda\ne 1/3$. Actually, as already
explained in the introduction, the regime of physical interest is
$\lambda>1$.

\subsection{Order analysis}

In order to expand the equations and to write down equations in each
order of gradient expansion, we need to know the orders of all relevant
variables. Therefore, we begin with the order analysis to determine
them.

Since we are interested in the spacetime which is not so much different
from the exact Friedmann universe, we suppose that
\begin{eqnarray}
 \partial_t \gamma_{ij} = O(\epsilon).
\end{eqnarray}
Substituting this into eq.(\ref{eq:evogama}), we obtain
\begin{eqnarray}
A^i_{\ j}=O(\epsilon).
\end{eqnarray}
Then, the constraint equation (\ref{eq:momentum-constraint}) implies
that 
\begin{eqnarray}
\partial_i K=O(\epsilon^2).
\end{eqnarray}
In other words, $K^{(0)}$ depends on $t$ only. This fact enables us to
define $a(t)$ by 
\begin{eqnarray}
3 \frac{\partial_t a(t)}{a(t)}= K^{(0)} (\equiv 3 H(t)).
\label{defa}
\end{eqnarray}
With this definition of $a(t)$, eq.~(\ref{eq:psi}) leads to 
\begin{eqnarray}
 \partial_t \zeta=O(\epsilon). 
\end{eqnarray}
In summary we have the following expansion. 
\begin{eqnarray}
 \zeta & = & \zeta^{(0)}(\vec{x}) +\epsilon \zeta^{(1)}(t,\vec{x})+ 
  \epsilon^2 \zeta^{(2)}(t,\vec{x})+\cdots , \\
 \gamma_{ij} & = & f_{ij}(\vec{x}) 
  +\epsilon \gamma_{ij}^{(1)}(t,\vec{x})+
  \epsilon^2 \gamma_{ij}^{(2)}(t,\vec{x})+\cdots ,\\
 K & = & 3 H(t)+\epsilon K^{(1)}(t,\vec{x})
  + \epsilon^2 K^{(2)}(t,\vec{x})+ \cdots, \\ 
 A^i_{\ j} & = & \epsilon A^{(1)\, i}_{\ \ \ \ \, j}(t,\vec{x}) +\epsilon^2
  A^{(2)\, i}_{\ \ \ \ \, j}(t,\vec{x}) + \cdots, 
\end{eqnarray}
where a quantity with the upper index $(n)$ is $n$-th order in
gradient expansion.

\subsection{Equations in each order}

We have found the orders of all physical quantities. Substituting this
into the evolution equations (\ref{eq:K}-\ref{eq:evogama}), we can
obtain the evolution equations in each order.  

In the zero-th order of gradient expansion we have
\begin{equation}
 (3\lambda-1)\left(\partial_t H + \frac{3}{2}H^2\right)
  = \Lambda. 
  \label{0thii}
\end{equation}
The first integral of this equation leads to
\begin{equation}
 3H^2 = \frac{2\Lambda}{3\lambda-1} + \frac{\tilde{C}}{a^3},
\end{equation}
where $\tilde{C}$ is an integration constant. The second term in the
right hand side of this equation is the ``dark matter as an integration
constant''~\cite{Mukohyama:2009mz}.

The $n$-th ($n\geq 1$) order equations are written as 
\begin{eqnarray}
 a^{-3}\partial_t \left(a^3 K^{(n)}\right)
 & = & 
 -\frac{1}{2}\sum_{p=1}^{n-1} K^{(p)}K^{(n-p)} 
 -\frac{3}{2(3\lambda-1)}\sum_{p=1}^{n-1}A^{(p)\, i}_{\ \ \ \ \, j}
 A^{(n-p)\, j}_{\qquad\ \ i}- \frac{Z^{(n)}}{3\lambda-1},
 \label{eq:n-K}\\
 a^{-3}\partial_t\left( a^3 A^{(n)\, i}_{\ \ \ \ \, j}\right) 
  & = &
  -\sum_{p=1}^{n-1}K^{(p)}A^{(n-p)\, i}_{\qquad\ \ j}
  +Z^{(n)\, i}_{\ \ \ \ \, j}-\frac{1}{3}Z^{(n)}\delta^i_{\ j},
  \label{eq:n-A}\\
 \partial_t \zeta^{(n)}
  & = & 
  \frac{1}{3} K^{(n)},\label{eq:n-zeta}\\
 \partial_t \gamma^{(n)}_{ij}
  & = &
  2 \sum_{p=0}^{n-1}\gamma^{(p)}_{ik}
  A^{(n-p)\, k}_{\qquad\ \ j},
  \label{eq:n-gamma}
\end{eqnarray}
In a similar way, from eq.(\ref{eq:momentum-constraint})  we obtain the
($n+1$)-th ($n\geq 1$) order momentum constraint equation as 
\begin{eqnarray}
 \partial_j A^{(n)\, j}_{\ \ \ \ \ i}
  +3 \sum_{p=1}^n A^{(p)\, j}_{\ \ \ \ \ i}\partial_j\zeta^{(n-p)}
  - \frac{1}{2}\sum_{p=1}^n\sum_{q=0}^{n-p}
  A^{(p)\, j}_{\ \ \ \ \ l}(\gamma^{-1})^{(q)\, lk}
  \partial_{i}\gamma^{(n-p-q)}_{jk}
  -\frac{1}{3}(3\lambda-1)\partial_i K^{(n)}
  =0,
\label{eq:n-const}
\end{eqnarray}
where $(\gamma^{-1})^{(n)\, ij}$ is the $n$-th order part of the
inverse of $\gamma_{ij}$, i.e. the inverse $(\gamma^{-1})^{ij}$ is
expanded as 
\begin{equation}
(\gamma^{-1})^{ij} = 
 f^{ij}
 + \epsilon (\gamma^{-1})^{(1)\, ij}
 + \epsilon^2 (\gamma^{-1})^{(2)\, ij}
 + \cdots, 
\end{equation}
where $f^{ij}=(\gamma^{-1})^{(0)\, ij}$ is the inverse of $f_{ij}$. It
is easy to show that $(\gamma^{-1})^{(n)\, ij}$ ($n\geq 1$) satisfies
the following differential equation. 
\begin{equation}
 \partial_t (\gamma^{-1})^{(n)\, ij} =
  -2\sum_{p=1}^n
  A^{(p)\, i}_{\quad\ \ k}
  (\gamma^{-1})^{(n-p)\, kj}. 
  \label{eqn:gamma-inv-nth-eq}
\end{equation}

There are some useful identities. The generalized Bianchi identity
(\ref{eq:DZ}) leads to
\begin{equation}
 \partial_j Z^{(n)\, j}_{\quad\ \ i}
  + 3\sum_{p=0}^n
  \left(Z^{(p)\, j}_{\quad\ \ i}
   -\frac{1}{3}Z^{(p)}\delta^j_{\ i}\right)\partial_j\zeta^{(n-p)}
  -\frac{1}{2}\sum_{p=0}^n\sum_{q=0}^{n-p}
  Z^{(p)\, j}_{\quad\ \ l}(\gamma^{-1})^{(q)\, lk}
  \partial_i\gamma^{(n-p-q)}_{jk} = 0, \label{eq:nth-DZ}
\end{equation}
where $Z^{(n)\, i}_{\quad\ \ j}$ and $Z^{(n)}$ are the $n$-th order 
parts of $Z^i_{\ j}$ and $Z$, respectively. The conditions 
$A^i_{\ i}=0$, $\partial_i\ln\det\gamma=0$, 
$\gamma_{ij}-\gamma_{ji}=0$, 
$\gamma_{ik}A^k_{\ j}-\gamma_{jk}A^k_{\ i}=0$ and 
$A^i_{\ j}-\gamma_{jk}A^k_{\ l}(\gamma^{-1})^{li}=0$ 
lead to the following identities. 
\begin{eqnarray}
& &
 A^{(n)\, i}_{\quad\ \ i} = 0, \quad
  \sum_{p=0}^n(\gamma^{-1})^{(p)\, jk}
  \partial_i\gamma^{(n-p)}_{jk} = 0, \quad
  \gamma^{(n)}_{ij}-\gamma^{(n)}_{ji}=0, \nonumber\\
 & & 
  \sum_{p=0}^{n-1}
  \left( \gamma^{(p)}_{ik}A^{(n-p)\, k}_{\qquad\ \ j}
   -\gamma^{(p)}_{jk}A^{(n-p)\, k}_{\qquad\ \ i}
       \right) = 0, \quad
 A^{(n)\, i}_{\quad\ \ j} - 
  \sum_{p=0}^{n-1}  \sum_{q=0}^{n-p-1}
  \gamma^{(p)}_{jk}A^{(n-p-q)\, k}_{\qquad\quad\quad l}
  (\gamma^{-1})^{(q)\, li}
   = 0. 
 \label{eqn:nth-identities}
\end{eqnarray}

\subsection{$O(\epsilon)$ solution}
\label{orderepsilon}

For the first order ($n=1$), eqs.(\ref{eq:n-K}-\ref{eq:n-gamma}) are
reduced to 
\begin{eqnarray}
\partial_t \left( a^3 K^{(1)} \right) & = & 0,
\label{eq:leading-Tr-EoM} \\
\partial_t \left( a^3 A^{(1)\, i}_{\ \ \ \ \, j} \right) & = & 0,
\label{eq:leading-Trless-EoM}\\
\partial_t \zeta^{(1)} & = & \frac{1}{3} K^{(1)},
\label{eq:leading-TrK}\\
\partial_t \gamma^{(1)}_{ij} & = &
 2 f_{ik} A^{(1)\, k}_{\ \ \ \ \, j}.
\label{eq:leading-TrlessK}
\end{eqnarray}
Note that $Z^{(1)i}_{\quad\ j}=0$. Integrating these equations, we
obtain 
\begin{eqnarray}
 K^{(1)} & = & \frac{C^{(1)}}{a(t)^3},
 \label{K1}\\
 A^{(1)\, i}_{\ \ \ \ \, j} & = & 
  \frac{C^{(1)\, i}_{\ \ \ \ \, j}}{a(t)^3},\\
 \zeta^{(1)} & = & 
  \frac{C^{(1)}}{3}
  \int^t_{t_{in}} \frac{dt'}{a^3(t')} + \zeta^{(1)}_{in},\\
 \label{zeta1}
  \gamma^{(1)}_{ij} & = & 
  2 f_{ik}C^{(1)\, k}_{\ \ \ \ \, j}
  \int^t_{t_{in}}\frac{dt'}{a^3(t')}
  + \gamma^{(1)}_{in\, ij},
\label{gamma1}
\end{eqnarray}
where the integration constants $C^{(1)}$, $C^{(1)\, i}_{\ \ \ \ \, j}$,
$\zeta^{(1)}_{in}$ and $\gamma^{(1)}_{in\, ij}$ depend on the spatial
coordinates $\vec{x}^i$ only, and satisfy
\begin{equation}
 C^{(1)\, i}_{\ \ \ \ \, i}=0,\quad
 f_{ik}C^{(1)\, k}_{\ \ \ \ \, j} =  f_{jk}C^{(1)\, k}_{\ \ \ \ \, i}.
\end{equation}
The two integration constants,
$\zeta^{(1)}_{in}$ and $\gamma^{(1)}_{in\, ij}$, can be absorbed into
the zero-th order counterparts, $\zeta^{(0)}_{in}$ and
$\gamma^{(0)}_{in\, ij}$. Thus, without loss of generality, we can set 
\begin{equation}
 \zeta^{(1)}_{in} = 0, \quad \gamma^{(1)}_{in\, ij}=0. 
\end{equation}
Finally, the momentum constraint equation (\ref{eq:n-const}) with $n=1$
leads to the following relation among the remaining integration
constants, $C^{(1)}$, $C^{(1)\, i}_{\ \ \ \ \, j}$, $\zeta^{(0)}$ and
$f_{ij}$. 
\begin{equation}
\partial_j C^{(1)\, j}_{\ \ \ \ \, i}
 + 3C^{(1)\, j}_{\ \ \ \ \, i}\partial_j\zeta^{(0)}
 - \frac{1}{2}C^{(1)\, j}_{\ \ \ \ \, l}f^{lk}\partial_i f_{jk}
 - \frac{1}{3}\left(3\lambda-1\right)\partial_iC^{(1)} = 0,
\label{const-1}
\end{equation}
where $f^{ij}$ is the inverse of $f_{ij}$.

\subsection{$O(\epsilon^2)$ solution}
\label{subsec:1storder}

With the first order solution obtained in the previous subsection, we 
can solve the second order equations. In the second order,
eqs.(\ref{eq:n-K}-\ref{eq:n-gamma}) become
\begin{eqnarray}
 a^{-3}\partial_t \left(a^3 K^{(2)}\right)
  & = & -\frac{1}{2}(K^{(1)})^2
  -\frac{3}{2((3\lambda-1))}
  A^{(1)\, i}_{\ \ \ \ \, j}A^{(1)\, j}_{\ \ \ \ \, i}
  -\frac{1}{2(3\lambda-1)}a^{-2}\tilde{R} ,\\
 a^{-3}\partial_t\left( a^3 A^{(2)\, i}_{\ \ \ \ \, j}\right)
  & = &
  -K^{(1)}A^{(1)\, i}_{\ \ \ \ \, j}
  -a^{-2}
  \left(\tilde{R}^i_{\ j}-\frac{1}{3}\tilde{R}\delta^i_{\ j}\right),\\ 
 \partial_t  \zeta^{(2)} & = & \frac{1}{3} K^{(2)} ,\\
 \partial_t \gamma^{(2)}_{ij} & = & 
  2 \left( f_{ik} A^{(2)\, k}_{\ \ \ \ \, j}
     +\gamma^{(1)}_{ik} A^{(1)\, k}_{\ \ \ \ \, j}\right),
\end{eqnarray}
where $\tilde{R}^i_{\ j}$ and $\tilde{R}$ are Ricci tensor and Ricci
scalar constructed from the 0-th order conformally-transformed metric 
$a^{-2}g^{(0)}_{ij}=e^{2\zeta^{(0)}} f_{ij}$, and we have used
the fact that $Z^{(2)\ i}_{\qquad j}=-a^{-2}
\left(\tilde{R}^i_{\ j}-\frac{1}{2}\tilde{R}\delta^i_{\ j}\right)$. 
By integrating these equations we obtain
\begin{eqnarray}
 K^{(2)} & = & -\frac{1}{2a^3(t)}
  \left\{
   \left[\left(C^{(1)}\right)^2
    + \frac{3}{3\lambda-1}
    C^{(1)\, i}_{\ \ \ \ \, j}C^{(1)\, j}_{\ \ \ \ \, i}
   \right]\int^t_{t_{in}}\frac{dt'}{a^3(t')} 
   + \frac{\tilde{R}}{3\lambda-1}
   \int^t_{t_{in}} a(t') dt' 
	\right\},\label{eq:2ndK}\\
 A^{(2)\, i}_{\ \ \ \ \, j} & = & 
  -\frac{1}{a^3(t)}
  \left\{
   C^{(1)}C^{(1)\, i}_{\ \ \ \ \, j}
   \int^t_{t_{in}}\frac{dt'}{a^3(t')} 
   + \left(\tilde R^{i}_{\ j}
      -\frac{1}{3}\tilde R \delta^i_{\ j}\right) 
   \int^t_{t_{in}} a(t') dt' 
	\right\},\\
 \zeta^{(2)} & = & 
  -\frac{1}{6}
  \left\{
   \left[\left(C^{(1)}\right)^2
  +\frac{3}{3\lambda-1}
  C^{(1)\, i}_{\ \ \ \ \, j}C^{(1)\, j}_{\ \ \ \ \, i}
  \right]
\int^t_{t_{in}}\frac{dt'}{a^3(t')}\int^{t'}_{t_{in}}\frac{dt''}{a^3(t'')}
 +\frac{\tilde{R}}{3\lambda-1} 
 \int^t_{t_{in}}\frac{dt'}{a^3(t')} 
 \int^{t'}_{t_{in}} a(t'') dt''\right\}, \label{zeta2}
 \\
 \gamma^{(2)}_{ij} & = & 2f_{ik} 
  \left[
   \left(
    2C^{(1)\, k}_{\ \ \ \ \, l}C^{(1)\, l}_{\ \ \ \ \, j}-
    C^{(1)}C^{(1)\, k}_{\ \ \ \ \, j} \right) 
   \int^t_{t_{in}}\frac{dt'}{a^3(t')}\int^{t'}_{t_{in}}\frac{dt''}{a^3(t'')}
   -\left(\tilde R^{ k}_{\  j}
     -\frac{1}{3}\tilde R \delta^k_{\ j}\right) 
   \int^t_{t_{in}}\frac{dt'}{a^3(t')}\int^{t'}_{t_{in}} a(t'') dt''
   \right], \label{eq:2ndgamma}
\end{eqnarray}
where we have set
\begin{equation}
 \left. K^{(2)}\right|_{t=t_{in}} =
 \left. A^{(2)\, i}_{\ \ \ \ \, j}\right|_{t=t_{in}} =
 \left. \zeta^{(2)}\right|_{t=t_{in}} =
 \left. \gamma^{(2)}_{ij}\right|_{t=t_{in}} = 0 
\end{equation}
by redefinition of $C^{(1)}$, $C^{(1)\, i}_{\ \ \ \ \, j}$, 
$\zeta^{(0)}$ and $f_{ij}$, respectively.

Provided that the redefined integration constants ($C^{(1)}$, 
$C^{(1)\, i}_{\ \ \ \ \, j}$, $\zeta^{(0)}$, $f_{ij}$) satisfy
(\ref{const-1}) up to $O(\epsilon^3)$, one can show that the solution
(\ref{eq:2ndK}-\ref{eq:2ndgamma}) automatically satisfies the
third-order momentum constraint equation 
\begin{eqnarray}
 & & \partial_j A^{(2)\, j}_{\ \ \ \ \, i} 
  + 3A^{(1)\, j}_{\ \ \ \ \, i}\partial_j\zeta^{(1)}
  + 3A^{(2)\, j}_{\ \ \ \ \, i}\partial_j\zeta^{(0)}
  \nonumber\\
 & &  \quad
  - \frac{1}{2}A^{(1)\, j}_{\ \ \ \ \, l}f^{lk}
  \partial_i\gamma^{(1)}_{jk}
  - \frac{1}{2}A^{(1)\, j}_{\ \ \ \ \, l}(\gamma^{-1})^{(1)\, lk}
  \partial_if_{jk}
  - \frac{1}{2}A^{(2)\, j}_{\ \ \ \ \, l}f^{lk}
  \partial_if_{jk} 
  - \frac{1}{3}\left(3\lambda-1\right)\partial_iK^{(2)}
  = 0. 
\end{eqnarray}
The proof is given in Appendix~\ref{proof}.

\subsection{$O(\epsilon^n)$ solution ($n\geq 2$)}

For general $n$ ($\geq 2$), the solution to
eqs.(\ref{eq:n-K}-\ref{eq:n-gamma}) is  
\begin{eqnarray}
 K^{(n)}
 & = & 
 \frac{1}{a^3(t)}\int^t_{t_{in}}dt' a^3(t')
 \left[
  -\frac{1}{2}\sum_{p=1}^{n-1} K^{(p)}K^{(n-p)} 
  -\frac{3}{2(3\lambda-1)}\sum_{p=1}^{n-1}A^{(p)\, i}_{\ \ \ \ \, j}
 A^{(n-p)\, j}_{\qquad\ \ i}- \frac{Z^{(n)}}{3\lambda-1}
 \right], \label{eq:nthK} \\
 A^{(n)\, i}_{\ \ \ \ \, j}
  & = &
 \frac{1}{a^3(t)}\int^t_{t_{in}}dt' a^3(t')
 \left[
  -\sum_{p=1}^{n-1}K^{(p)}A^{(n-p)\, i}_{\qquad\ \ j}
  +Z^{(n)\, i}_{\ \ \ \ \, j}-\frac{1}{3}Z^{(n)}\delta^i_{\ j}
  \right], \label{eq:nthA} \\
 \zeta^{(n)}
  & = & 
  \frac{1}{3} \int^t_{t_{in}}dt' K^{(n)},\\
 \gamma^{(n)}_{ij}
  & = &
  2 \int^t_{t_{in}}dt' \sum_{p=0}^{n-1}\gamma^{(p)}_{ik}
  A^{(n-p)\, k}_{\qquad\ \ j}, \label{eq:nthgamma}
\end{eqnarray}
where we have set
\begin{equation}
 \left. K^{(n)}\right|_{t=t_{in}} =
 \left. A^{(n)\, i}_{\ \ \ \ \, j}\right|_{t=t_{in}} = 
 \left. \zeta^{(n)}\right|_{t=t_{in}} =
 \left. \gamma^{(n)}_{ij}\right|_{t=t_{in}} = 0
\end{equation}
by redefinition of $C^{(1)}$, $C^{(1)\, i}_{\ \ \ \ \, j}$,
$\zeta^{(0)}$ and $f_{ij}$, respectively. Note that, in
subsection~\ref{subsec:1storder}, we have already set 
\begin{equation}
 \left. \zeta^{(1)}\right|_{t=t_{in}} =
 \left. \gamma^{(1)}_{ij}\right|_{t=t_{in}} = 0
\end{equation}
by redefinition of $\zeta^{(0)}$ and $f_{ij}$, respectively.

The initial condition for $\gamma^{(n)}_{ij}$ ($n\geq 1$) implies that 
$\left.\gamma_{ij}\right|_{t=t_{in}}=f_{ij}$, that 
$\left.(\gamma^{-1})^{ij}\right|_{t=t_{in}}=f^{ij}$ and that 
$\left.(\gamma^{-1})^{(n)\, ij}\right|_{t=t_{in}}=0$ ($n\geq 1$). 
Therefore, for $n\geq 1$, the solution to (\ref{eqn:gamma-inv-nth-eq})
is 
\begin{equation}
 (\gamma^{-1})^{(n)\, ij} =
  -2\int_{t_{in}}^t dt'\sum_{p=1}^n
  A^{(p)\, i}_{\quad\ \ k}
  (\gamma^{-1})^{(n-p)\, kj}. 
\end{equation}

As shown in Appendix~\ref{proof}, the solution
(\ref{eq:nthK}-\ref{eq:nthgamma}) automatically satisfies the
($n+1$)-th order momentum constraint equation (\ref{eq:n-const}),
provided that the redefined integration constants ($C^{(1)}$, 
$C^{(1)\, i}_{\ \ \ \ \, j}$, $\zeta^{(0)}$, $f_{ij}$) satisfy
(\ref{const-1}) up to $O(\epsilon^{n+1})$.

\subsection{Number of physical degrees of freedom}

Our solution involves functions $\zeta^{(0)}(\vec{x})$, 
$f_{ij}(\vec{x})$, $C^{(1)}(\vec{x})$ and 
$C^{(1)\, i}_{\ \ \ \ \, j}(\vec{x})$ of spatial coordinates as
integration `constant`. They are subject to the constraint
(\ref{const-1}). Also, as stated just after (\ref{eq:gaugefix}), our
gauge fixing condition (\ref{eq:gaugefix}) leaves time-independent
spatial diffeomorphism as residual gauge freedom. Therefore, the number
of physical degrees of freedom included in each integration `constant` is 
%
\begin{eqnarray}
 \zeta^{(0)}(\vec{x}) & \cdots & 
  1 \mbox{ scalar growing mode }
  = 1 \mbox{ component }, \nonumber\\
 f_{ij}(\vec{x}) & \cdots & 
  2 \mbox{ tensor growing modes }
  = 5 \mbox{ components } - 3 \mbox{ gauge }, \nonumber\\
 C^{(0)}(\vec{x}) & \cdots & 
  1 \mbox{ scalar decaying mode }
  = 1 \mbox{ component }, \nonumber\\
 C^{(1)\, i}_{\ \ \ \ \, j}(\vec{x}) & \cdots & 
  2 \mbox{ tensor decaying modes }
  = 5 \mbox{ components } - 3 \mbox{ constraints }. 
\end{eqnarray}
This is consistent with the fact that the Ho\v{r}ava-Lifshitz gravity
includes not only a tensor graviton ($2$ propagating degrees of freedom)
but also a scalar graviton ($1$ propagating degree of freedom).

\section{Conserved curvature perturbation}
\label{CP}

In this section we propose a definition of nonlinear curvature
perturbation and show that it is conserved up to the first order in the
gradient expansion. This statement holds for any values of
$\lambda$. Because of the conservation, this quantity is expected to be
useful for the analysis of superhorizon evolution of nonlinear
perturbation.

\subsection{Definition}

In general relativity, it is known that the curvature perturbation in
the uniform density slice conserves up to the first order in the gradient
expansion~\cite{Salopek:1990jq}. Motivated by this fact, we
define the quantity ${\cal R}$ in Ho\v{r}ava-Lifshitz gravity by 
\begin{equation}
 {\cal R}(t,\vec{x}) \equiv \zeta(\tilde{t},\vec{x}) 
  + \ln\left[\frac{a(\tilde{t})}{a(t)} \right], 
\label{Rdef}
\end{equation}
where $\tilde{t}(t,\vec{x})=t+O(\epsilon)$ is the solution to 
\begin{equation}
 \rho_{dm}^{(0)}(\tilde{t}) + 
 \delta\rho_{dm}(\tilde{t},\vec{x})  
  = \rho_{dm}^{(0)}(t),
\end{equation}
and 
\begin{equation}
 \rho_{dm}^{(0)}(t) \equiv 
  3M_{Pl}^2H^2-\frac{2M_{Pl}^2}{3\lambda-1}\Lambda, \quad
  \delta\rho_{dm}(t,\vec{x})
  \equiv \frac{M_{Pl}^2}{2}
  \left[ R + \frac{2}{3}(K^2-9H^2) - A^i_jA^j_i\right]. 
\end{equation}

In the following we shall show that ${\cal R}$ is indeed conserved up to
the first order in the gradient expansion by explicitly calculating it.

\subsection{Concrete expression and conservation up to $O(\epsilon)$}

According to the gradient expansion, we expand $\delta\rho_{dm}$,
$\tilde{t}$ and ${\cal R}$ as
\begin{equation}
 \delta\rho_{dm} = \sum_{n=1}^{\infty}\rho_{dm}^{(n)}, \quad
 \tilde{t} = t + \sum_{n=1}^{\infty}\tilde{t}^{(n)}, \quad
 {\cal R} = \sum_{n=0}^{\infty}{\cal R}^{(n)}, 
\end{equation}
where a quantity with the superscript $(n)$ is of order
$O(\epsilon^n)$. 

At the order $O(\epsilon^0)$, we obtain
\begin{equation}
 {\cal R}^{(0)} = \zeta^{(0)}(\vec{x}),
 \end{equation}
and this is constant in time. 

At the order $O(\epsilon)$, we obtain
\begin{equation}
 \rho_{dm}^{(1)} = 2M_{Pl}^2 H K^{(1)}, \quad
  \tilde{t}^{(1)} = -\frac{\rho_{dm}^{(1)}}{\partial_t \rho^{(0)}}
  = -\frac{K^{(1)}}{3\partial_t H}, \quad
  {\cal R}^{(1)} = \zeta^{(1)} + H\tilde{t}^{(1)}
  = \frac{C^{(1)}}{3}
  \left[\int^t_{t_{in}}\frac{dt'}{a^3(t')}- \frac{H}{a^3\partial_t H}
  \right].
\end{equation}
Thus, the time derivative of ${\cal R}^{(1)}$ is shown to vanish as
\begin{equation}
 \partial_t {\cal R}^{(1)} = \frac{C^{(1)}H}{3a^3(\partial_t H)^2}
  (\partial_t^2 H+3H\partial_t H) = 0,
\end{equation}
where we have used the time derivative of (\ref{0thii}) to show the last
equality.

We therefore conclude that ${\cal R}$ defined in (\ref{Rdef}) is
conserved up to the first order in the gradient expansion. This
statement holds for any values of $\lambda$.

\section{Summary and Discussion}
\label{sec:summary}

We have performed a fully nonlinear analysis of superhorizon
perturbation in Ho\v{r}ava-Lifshitz gravity by using the gradient
expansion technique. In \S~\ref{solutions} we have presented a concrete
expression for the solution of gravity equations up to the second order
in the gradient expansion. We have also proven that the solution can be 
extended to any order of the gradient expansion, by showing that the
solution to the dynamical equation satisfies the constraint equation at
each order.

Based on the result, in \S~\ref{CP} we have proposed a definition of
nonlinear curvature perturbation ${\cal R}$ in Ho\v{r}ava-Lifshitz
gravity and have shown that it is conserved up to the first order in the
gradient expansion.

It is known that, in the limit $\lambda\to 1$, the scalar graviton gets
strongly coupled and the usual metric perturbation breaks down in the
scalar graviton sector~\cite{Charmousis:2009tc}. Here, we stress that the
breakdown of the perturbative expansion does not necessarily lead to a
loss of predictability since all coefficients of infinite number of terms in the
perturbative expansion can be written in terms of finite parameters in
the action if the theory is renormalizable. Indeed, for spherically
symmetric, static, vacuum configurations, one can perform fully
nonlinear analysis to show that the limit $\lambda\to 1$ is continuous
and the general relativity is recovered in the
limit~\cite{Mukohyama:2010xz}. This result may be considered as an
analogue of Vainshtein effect and suggests the possibility that the
scalar graviton may safely be decoupled from the rest of the world,
i.e. the tensor graviton and the matter sector, in the limit.

The result of the present paper is based on the fully nonlinear analysis
and may be considered as yet another example of the analogue of
Vainshtein effect. Up to any order of the gradient expansion, the
equations of motion and their solutions are manifestly regular in the
limit $\lambda\to 1$. The solutions reduce to those in general
relativity coupled to dark matter in the limit at low energy.

In the present paper we have concentrated on the pure gravity system in
the projectable Ho\v{r}ava-Lifshitz theory. Because of the existence of
the scalar graviton, this simple system is still rich enough as a
testing ground for the analogue of Vainshtein effect. Indeed, the so
called ``dark matter as integration constant''~\cite{Mukohyama:2009mz}
drives non-trivial cosmological dynamics in this system, and thus the
nonlinear analysis presented in the present paper provides a convincing
evidence for the analogue of Vainshtein effect. It is certainly
interesting and important to extend the nonlinear analysis to more
general situations with matter contents~\cite{GMW}.

\begin{acknowledgments} 
 K.I. acknowledges supports by the Grant-in-Aid for Scientific Research 
 (A) No. 21244033. Part of this work was done during S.M.'s
 participation in YITP molecule-type workshop (T-10-05): Cosmological
 Perturbation and Cosmic Microwave Background. He thanks YITP for
 stimulating atmosphere and warm hospitality. The work of S.M. is
 supported by Grant-in-Aid for Scientific Research 17740134, 19GS0219,
 21111006, 21540278, and by World Premier International Research Center
 Initiative (WPI Initiative). The authors are also supported by
 Japan-Russia Research Cooperative Program. 
\end{acknowledgments}

\appendix
\section{($n+1$)-th order momentum constraint ($n\geq 2$)}
\label{proof}

In this appendix, by induction we prove that the $n$-th order solution 
(\ref{eq:nthK}-\ref{eq:nthgamma}) satisfies the ($n+1$)-th order
momentum constraint equation (\ref{eq:n-const}) for $n\geq 2$. 

The basic logic of the proof is to rewrite the left hand side of the 
($n+1$)-th order constraint (\ref{eq:n-const}) as a linear combination
of lower order constraints by using the explicit solution
(\ref{eq:nthK}-\ref{eq:nthgamma}). For this purpose we shall use the
generalized Bianchi identity (\ref{eq:nth-DZ}) and other identities
(\ref{eqn:nth-identities}). We also use the following identity for
functions $f(t)$ and $g(t)$ satisfying
$a^3(t_{in})f(t_{in})g(t_{in})=0$: 
\begin{equation}
 f(t)g(t) = \frac{1}{a^3(t)}\int_{t_{in}}^t dt'a^3(t')
  \left[a(t')^{-3}\partial_{t'}(a^3(t')f(t'))\cdot g(t')
   + f(t')\cdot \partial_{t'}g(t')\right]. \label{eqn:integration-by-part}
\end{equation}

By applying the identity (\ref{eqn:integration-by-part}) to
$(f(t),g(t))=(A^{(p)\, j}_{\ \ \ \ \ i},\partial_j\zeta^{(n-p)})$ and 
$(f(t),g(t))=(A^{(p)\, j}_{\ \ \ \ \ l},
(\gamma^{-1})^{(q)\, lk}\partial_{i}\gamma^{(n-p-q)}_{jk})$, 
the left hand side of the ($n+1$)-th order momentum constraint equation
(\ref{eq:n-const}) is rewritten as
\begin{eqnarray}
 {\cal C}^{(n+1)}_i & \equiv & 
  \partial_j A^{(n)\, j}_{\ \ \ \ \ i}
  +3 \sum_{p=1}^n A^{(p)\, j}_{\ \ \ \ \ i}\partial_j\zeta^{(n-p)}
  - \frac{1}{2}\sum_{p=1}^n\sum_{q=0}^{n-p}
  A^{(p)\, j}_{\ \ \ \ \ l}(\gamma^{-1})^{(q)\, lk}
  \partial_{i}\gamma^{(n-p-q)}_{jk}
  -\frac{1}{3}(3\lambda-1)\partial_i K^{(n)} \nonumber\\
  &=& 
   \partial_j A^{(n)\, j}_{\ \ \ \ \ i}
   + \frac{1}{a^3(t)}\int_{t_{in}}^t dt' a^3(t')
   \left\{
    3\sum_{p=1}^n\left[
      a^{-3}\partial_{t'}
      \left( a^3A^{(p)\, j}_{\ \ \ \ \ i}\right)
      \partial_j\zeta^{(n-p)}
      + A^{(p)\, j}_{\ \ \ \ \ i}
      \partial_j\left(\partial_{t'}\zeta^{(n-p)}\right)
     \right]
    \right. \nonumber\\
 & & 
  - \frac{1}{2}\sum_{p=1}^n\sum_{q=0}^{n-p}
  \left[ a^{-3}\partial_{t'}
   \left( a^3A^{(p)\, j}_{\ \ \ \ \ l}\right)
   (\gamma^{-1})^{(q)\, lk}
   \partial_{i}\gamma^{(n-p-q)}_{jk}
   + A^{(p)\, j}_{\ \ \ \ \ l}
   \partial_{t'}\left((\gamma^{-1})^{(q)\, lk}\right)
  \partial_{i}\gamma^{(n-p-q)}_{jk}\right.
  \nonumber\\
 & & \left.\left.
   + A^{(p)\, j}_{\ \ \ \ \ l}
   (\gamma^{-1})^{(q)\, lk}
  \partial_{i}\left(\partial_{t'}\gamma^{(n-p-q)}_{jk}\right)
  \right]\right\}
  -\frac{1}{3}(3\lambda-1)\partial_i K^{(n)}. \nonumber\\
\end{eqnarray}
Using (\ref{eq:nthK}-\ref{eq:nthA}), (\ref{eq:n-zeta}-\ref{eq:n-gamma})
and (\ref{eqn:gamma-inv-nth-eq}), this is further rewritten as 
\begin{eqnarray}
 {\cal C}^{(n+1)}_i 
  &=& 
  \frac{1}{a^3(t)}\int_{t_{in}}^t dt' a^3(t')
  \left\{
   \partial_j
   \left( -\sum_{p=1}^{n-1}K^{(p)}A^{(n-p)\, j}_{\qquad\ \ i}
   \right)
    \right. \nonumber\\
 & & 
   +3\left[
      \sum_{p=2}^n
      \left(-\sum_{q=1}^{p-1}K^{(q)}A^{(p-q)\, j}_{\qquad\ \ i}\right)
      \partial_j\zeta^{(n-p)}
      + \sum_{p=1}^{n-1}A^{(p)\, j}_{\ \ \ \ \ i}
      \partial_j\left(\frac{1}{3}K^{(n-p)}\right)
	    \right]
   \nonumber\\
 & & 
  - \frac{1}{2}\sum_{p=1}^n\sum_{q=0}^{n-p}
  \left[ 
   \left(-\sum_{r=1}^{p-1}K^{(r)}A^{(p-r)\, j}_{\qquad\ \ l}\right)
   (\gamma^{-1})^{(q)\, lk}
   \partial_{i}\gamma^{(n-p-q)}_{jk}
   \right.\nonumber\\
 & &
  \left.
   + A^{(p)\, j}_{\ \ \ \ \ l}
   \left(-2\sum_{r=1}^q
  A^{(r)\, l}_{\quad\ \ m}
  (\gamma^{-1})^{(q-r)\, mk}\right)
  \partial_{i}\gamma^{(n-p-q)}_{jk}\right.
  \nonumber\\
 & & \left.
   + A^{(p)\, j}_{\ \ \ \ \ l}
   (\gamma^{-1})^{(q)\, lk}
   \partial_{i}
   \left(
    2 \sum_{r=1}^{n-p-q-1}\gamma^{(r)}_{jm}
    A^{(n-p-q-r)\, m}_{\qquad\qquad\quad k}
	   \right)\right]
  \nonumber\\
 & & 
 -\frac{1}{6}(3\lambda-1)\partial_i 
 \left(
 -\sum_{p=1}^{n-1} K^{(p)}K^{(n-p)} 
 \right)
 -\frac{1}{2}\partial_i
 \left(
  -\sum_{p=1}^{n-1}A^{(p)\, j}_{\ \ \ \ \, k}
  A^{(n-p)\, k}_{\qquad\ \ j}
       \right)
  \nonumber\\
 & & \left.
 + \frac{1}{6}\sum_{p=1}^nZ^{(p)}
 \sum_{q=0}^{n-p}
  (\gamma^{-1})^{(q)\, jk}
  \partial_{i}\gamma^{(n-p-q)}_{jk}
     \right\},
\end{eqnarray}
where we have used the generalized Bianchi identity
(\ref{eq:nth-DZ}). By using the identities (\ref{eqn:nth-identities}) we
finally obtain 
\begin{equation}
 {\cal C}^{(n+1)}_i = 
  -\frac{1}{a^3(t)}\int_{t_{in}}^t dt' a^3(t')
  \sum_{p=1}^{n-1}K^{(n-p)}{\cal C}^{(p+1)}_i. 
\end{equation}
Since we already know that ${\cal C}^{(2)}_i=0$ under the condition 
(\ref{const-1}), this is enough to prove ${\cal C}^{(n+1)}=0$ for 
$n\geq 2$.

\end{document}